\documentclass[aps,twocolumn,groupedaddress,prl,bibnotes,showpacs]{revtex4}
\usepackage{graphicx}
\begin{document}


\title[]{Pure spin decoherence in quantum dots.}

\author{Y.~B.~Lyanda-Geller}
\email[]{yuli@purdue.edu}
\affiliation{
Department of Physics and Birck Nanotechnology Center, Purdue University, West Lafayette, Indiana
47907 USA}

\date{\today, ---= draft: \jobname.tex =---}

\begin{abstract}
We uncover two microscopic physical settings with significant pure spin decoherence. First, for quantum dots (QD) electrostatically 
confined in two-dimensional hole gas, decoherence comes from qubit spin-orbit (SO) coupling to phonons, whose decay due to free charge carriers in contacts is described by Ohmic weighted phonon spectral function. We derive significant SO interactions affecting holes with origin and symmetry distinct from that of conventionally considered Dresselhaus and Rashba terms. In the second setting of electron or hole QDs coupled to a linear chain of atoms, decoherence is due to spin-dependent coupling to phonons, whose decay due to scattering off the free ends of a chain is described by the weighted phonon spectral function inverse proportional to frequency. 
The decoherence rate in both settings is linear in temperature.
\end{abstract}

\maketitle

When spin population is conserved, quantum coherent properties are defined by pure spin decoherence\cite{Slichter}.  In optics, there is a related phenomenon of zero phonon spectral lines\cite{Krivoglaz,Skinner,Gupalov}.  For the field of quantum computing, understanding mechanisms of the pure spin decoherence is vital. In  quantum dot (QD) qubits, the task is to uncover qubit-phonon interactions and processes in phonon system that lead to the effect. 
Current understanding of QDs does not include pure spin decoherence due to spin-orbit (SO) interactions and
phonons as an effective mechanism, and finding to the contrary is important.

\begin{figure}[t]
\vspace{-11mm}
\includegraphics[scale=0.2]{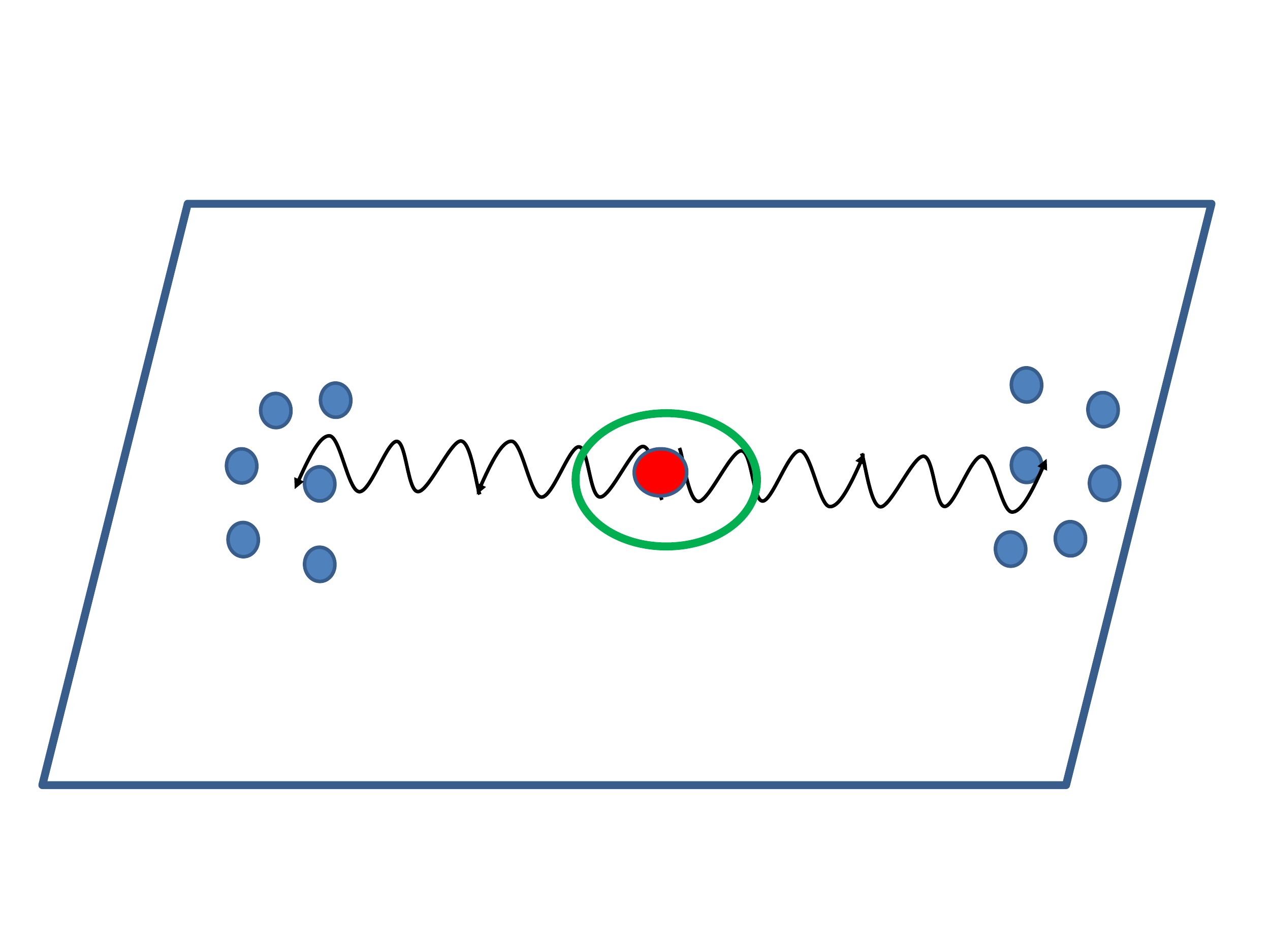}
\vspace{-9mm}
\caption{QD in a 2D hole gas. Pure spin decoherence is due to 
interaction of a localized hole spin with phonons that in turn decay due to 
interaction with delocalized holes in contacts. 
}
\vspace{-1mm}
\label{Fig1}
\end{figure}

\begin{figure}[t]
\vspace{-5mm}
\includegraphics[scale=0.2]{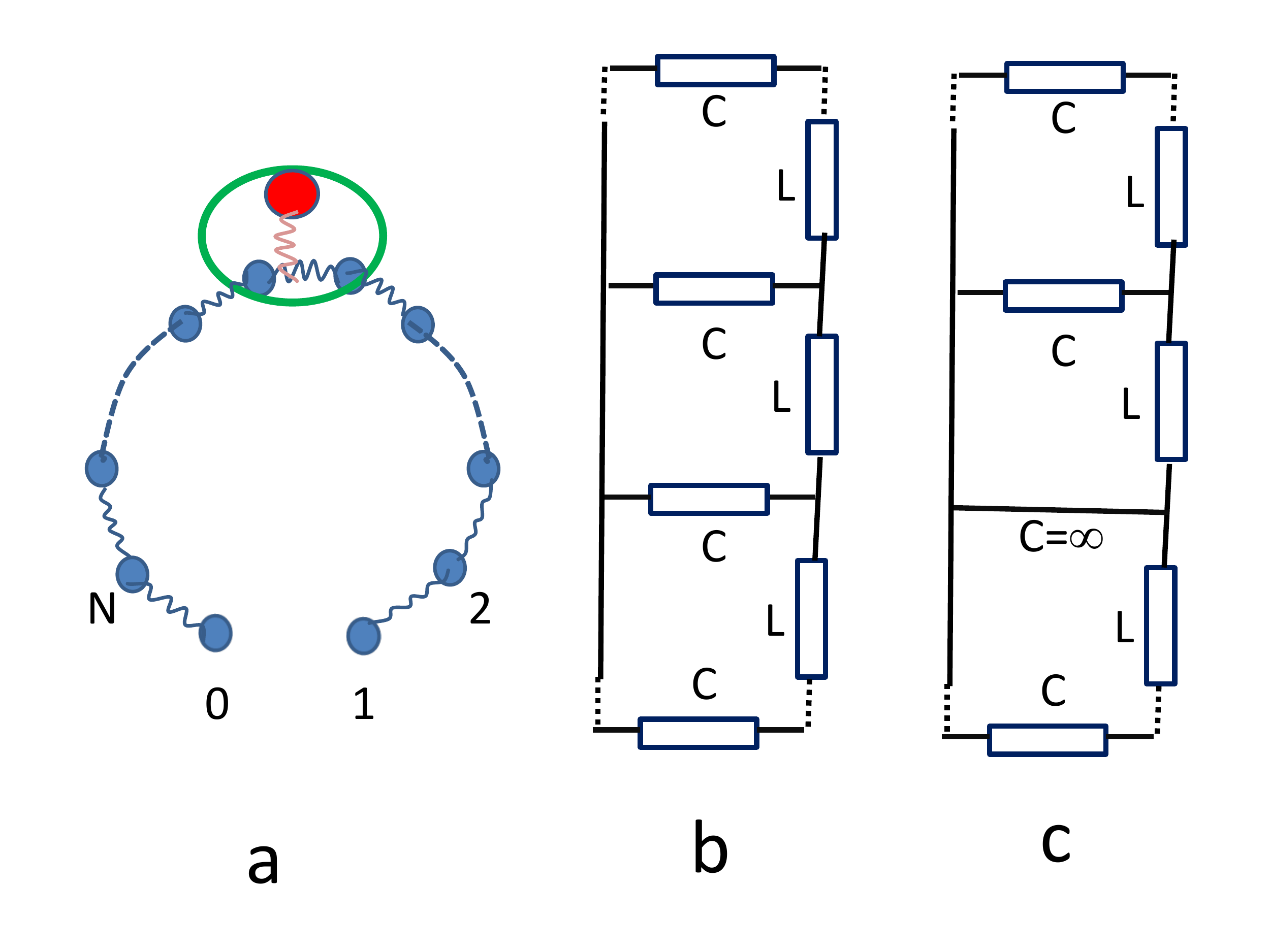}
\vspace{-6mm}
\caption{a. QD coupled to a chain of atoms. Pure decoherence is due to spin coupling to phonons that scatter of the free ends (0,1) of a chain. b. Inductance-capacitance circuit equivalent to an ideal chain. c. Circuit for a chain with missing spring. }  
\vspace{-5mm}
\label{Fig2}
\end{figure}

In this letter I show that pure spin decoherence is crucial for 
QDs. In particular, I consider QD qubits with charge carrier holes, in which spin 
relaxation is suppressed, and nuclear spin effects are weak compared to electron QDs. I derive spin-phonon 
interactions in hole QDs giving sizable pure spin decoherence.  I show that steps can be taken to suppress it. 
I find that pure spin decoherence in QDs can be related to realistic microscopic 
mechanisms of phonon decay. It arises for a hole localized 
in a QD electrostatically confined in the two-dimensional (2D) gas, when the hole interacts 
with phonons, which in turn decay due to interaction with free holes present in the system, Fig.\ref{Fig1}. 
This is a microscopic realization of models with Ohmic phonon weighted spectral density. The other setting for pure decoherence, Fig.\ref{Fig2}, is electronic or hole QD coupled to 1D atomic chain, 
due to qubit-phonon coupling and phonon scatttering of the free ends of the chain, which results in a weighted spectral density 
inverse proportional to frequency. 

I derive the effective Hamiltonian for QDs 
confined to 2D hole systems, which also describes rings, wires and point contacts.  The problem was extensively 
addressed in recent years \cite{Arovas,Winkler,finestructure,Losssr,Ulloa,Lin,Schlieman,bernevig,Lossrelax,Zuelicke,chesi,kimSiGe}, but an important effect was missed. 

In the ground state of the confined hole systems, like heterostructures 
in iii-v or Si/Ge structures grown along 001 crystallographic direction, the effective mass $m_h$
in the growth direction $z$ is a heavy hole mass. This state 
is attractive for qubits \cite{Lossrelax}, because it is mostly characterized by the angular 
momentum projection $L_z=\pm 3/2$. Since the matrix elements of 
spin-flipping interactions due to $J_x$ and $J_y$ between states +3/2 and -3/2 vanish, decoherence of hole QDs is reduced 
\cite{linear,YLGHSC} compared to electronic QDs. 
Understanding hole qubits requires knowledge of the effective Hamiltonian projected on the hole ground state doublet that must account for 
mixing of $\pm 3/2$ states to the states 
with light effective mass $m_l$ in $z$-direction ( $L_z=\pm 1/2$). 

To date, SO interactions for 2D and QD holes
have been treated similar to those for electrons, and 
limited to modified Dresselhaus and Rashba terms, which do not give pure 
spin decoherence. Here we show that SO interactions distinct from the Dresselhaus and Rashba 
terms affect 2D and QD holes. They arise even if bulk crystal is centrosymmetric (no Dresselhaus term \cite{Dresselhaus}), and quantum wells (QW) are symmetric (no Rashba term \cite{rashba1}).
Their origin is the hole Luttinger Hamiltonian \cite{Luttinger}
\begin{eqnarray}
&{\cal H} =  \frac{1}{2m_0}\Big( (\gamma_1+\frac{5}{2}\gamma_2) p^2 I -2\gamma_2({\mathbf J}\cdot{\mathbf p})^2 \nonumber\\
&-4(\gamma_3-\gamma_2) \sum_{i>j}p_ip_j[J_iJ_j]\Big),
\label{H1}
\end{eqnarray}
where $J_i$ are the 4$\times $4matrices of 
the $i-$component of $J=3/2$ operator, 
[AB]=(AB+BA)/2, $\gamma_i$ are Luttintger constants, $m_0$ is the free electron mass. 
Infinite rectangular QW, in which Hamiltonian Eq.(\ref{H1}) 
is supplemented by the boundary conditions giving vanishing wavefunctions at $z=\pm d/2$, allows analytical solution 
\cite{nedoresov,khaetsky,portnoi}. Wavefunctions  $\Psi_s(z)$, $\Psi_a(z)$ are 
symmetric and antisymmetric with respect to reflection about the plane $z=0$, and
in the basis of Bloch functions $(u^{3/2}, u^{1/2}, u^{-1/2}, u^{-3/2})$ describing bulk holes with 
$J=3/2$, are given by $\Psi_{\pm}(z,{\mathbf r})
=a^{\mathbf k, z}_{\pm}\exp{(i{\mathbf k}\cdot{\mathbf r})}$, where $\mathbf{r}$ is the in-plane coordinate, and
\begin{eqnarray}
a^{\mathbf k,z}_-=\left(
\begin{array}{c}
iA_3 S_ze^{-3i\phi_k}\\ A_2 C_ze^{-2i\phi_k}\\iA_1 S_z e^{-i\phi_k}\\ A_0 C_z
\end{array}
\right).
\hspace{2mm}
a^{\mathbf k,z}_+=\left(
\begin{array}{c}
A_0 C_z\\ -iA_1 S_z e^{i\phi_k}\\A_2 C_ze^{2i\phi_k}\\-iA_3 S_ze^{3i\phi_k}  
\end{array}
\right),
\label{wf}
\end{eqnarray}
$\mathbf{k}$ is the in-plane momentum, $\phi_k$ is the angle 
between $\mathbf{k}$ and $x-$direction.
Here $S_z=\sin{q_hz}-(s_h/s_l)\sin{q_lz}$, $C_z=\cos{q_hz}-(c_h/c_l)\cos{q_lz}$,
are antisymmetric and symmetric standing waves with respect to $z=0$, $s_h=\sin{q_hd/2}$, 
$s_l=\sin{q_ld/2}$, $c_h=\cos{q_hd/2}$ 
$c_l=\cos{q_ld/2}$, and $d$ is the QW width.  For 
energies of in-plane motion $\epsilon_{\mathbf{k}}\ll \epsilon_q=\hbar^2q_h^2/2m_h$, the wavevectors $q_h=\pi/d$ and $q_l= \sqrt{\nu}q_h$, where
$\nu=m_l/m_h$. 
Thus, QW holes in contrast to QW electrons are described by 
two standing waves, illustrating
mutual transformation of heavy and light holes upon reflection from the barrier.
In the spherical 
approximation $\gamma_2=\gamma_3$ and at small in-plane energies, the normalization factor $A_0=\Big(\frac{d}{2}+\frac{4\nu}{3}I_s\Big)^{-1/2}$, $I_s=(1+\frac{1}{s_l^2})\frac{d}{2}+\frac{c_l}{s_l}\frac{(q_h^2+3q_l^2)}{q_l(q_l^2-q_h^2)}$, and
amplitudes $A_1=A_0\frac{\sqrt{3} k}{2q_h}$, $A_2= A_0\frac{\sqrt{3} k^2}{4q_h^2}$ and 
$A_3= 3A_0k^3/(8\nu q_h^3)$.
 
Taking the projection \cite{note} of $\hat{V}(\mathbf{r},z,t)$ that includes potential confining holes to QDs and 
potential due to long-range hole-phonon interactions \cite{longrange,pikus}
onto the ground state given by Eqs.(\ref{wf}), in the lowest order in $\mathbf{k}$ 
we derive
\begin{eqnarray}
&{\cal H}= \frac{\hbar^2k^2}{2m^*}+ \alpha_0 {\tilde V}_0({\mathbf r},t) +\alpha_1{\tilde V}_1({\mathbf r},t)
-i \alpha_2{\mathbf k}\cdot\nabla_{\mathbf r}{\tilde V_1}({\mathbf r},t)\nonumber\\
&\alpha_2\sigma_z[{\mathbf k}\times 
\nabla_{\mathbf r}{\tilde V_1}({\mathbf r},t)]_z,
\label{holegen}
\end{eqnarray}
where $m^*$ is the effective mass given by
\begin{equation}
\frac{m_0}{m^*}=\frac{\gamma_2(\gamma_1+\gamma_2)-3\gamma^2_3}{\gamma_2}+
\frac{3\gamma^2_3(\gamma_1^2-4\gamma_2^2)^{1/2}}{\gamma_2^2}f(\theta),
\label{in-plane}
\end{equation}
where $m_0$ is the free electron mass, $f(\theta)=\frac{1+\cos{\theta}}{\pi \sin{\theta}}$,
  $\theta=\pi\sqrt{ \frac{\gamma_1-2\gamma_2}{\gamma_1+2\gamma_2}}$,
$\alpha_0=A_0^2d$, $\alpha_1= \frac{3A_0^2k^2d^3}{4\pi^2}$, 
$\alpha_2=\frac{3A_0^2d^3}{4\pi^2}$ and  projection potentials
${\tilde V}_0({\mathbf r},t)= \frac{1}{d}\int_{-d/2}^{d/2} V(({\mathbf r},z,t)C^2(z)dz$, 
${\tilde V}_1({\mathbf r},t)= \frac{1}{d}\int_{-d/2}^{d/2} V(({\mathbf r},z,t)S^2(z)dz$.
The SO terms in (\ref{holegen}) give no spin-flips, and have $U(1)$ symmetry,  partially breaking spin rotation symmetry \cite{LGM}.  
Spin flip terms from projection of $\hat{V}(\mathbf{r},z,t)$ onto $\Psi_{\pm}(z,{\mathbf r})$ contain at least the parameter $(kd)^3$ and 
are smaller than SO term in (\ref{holegen}) at small in-plane energies. 
We note that  for 2D holes, the Rashba term is $\propto k^3$, and the Dresselhaus terms 
include both cubic and linear in $\mathbf{k}$ contributions 
\cite{Sherman}. Thus, the SO
term in (\ref{holegen})
missed in all previous work on hole QWs and QDs has profound consequences: in Si-Ge systems, it dominates; in iii-v systems, 
it changes the symmetry of SO interactions. 
This term 
is of sizable strength, defined by $kd$ and averaging over $z$-direction, 
and is {\it not} related to admixture of distant bands like electron SO terms. 

We now discuss pure decoherence in small QDs, with sizes less than hole mean free 
path. We take parabolic projection potential $\alpha_0\tilde{V}_0+\alpha_1\tilde{V}_1=m^{*}(\omega_1^2x^2+\omega_2^2y^2)/2$ 
to model 
in-plane spatial confinement of holes. The linear in $\mathbf{k}$ and $\tilde{V_1}$ term gives negligible mixture of orbital levels and is omitted. 
The relevant SO interaction is the last term in (\ref{holegen}) due to hole-phonon interaction $V(z,\mathbf{r},t)$. We will
refer to it as spin-phonon interaction
\begin{equation}
{\cal H}_{sph}=\mathbf{\sigma}_z\sum_{\mathbf{q},j}V_{\mathbf{q},j}(c^{\dagger}_{\mathbf{q},j}-c_{\mathbf{-q},j})=\hbar\mathbf{\sigma}_z\Omega_z,
\label{spinphonon}
\end{equation}
where  $V_{\mathbf{q},j}$ is the Fourier component of the amplitude of spin-phonon interaction, $j$ denote the phonon mode polarization, 
and $c^{\dagger}_{\mathbf{q},j}$, $c_{\mathbf{q},j}$ are 
phonon creation/annihilation operators, which define lattice displacement operator 
$\hat{\mathbf{ R}}_{np}= \sum_{\mathbf{q},j}\mathbf{u}_{np}(\mathbf{q},j)(\hat{c}^{\dagger}_{\mathbf{q},j}+\hat{c}_{-\mathbf{q},j})$, where $n$  labels the crystalline unit cell, $p$ are the atoms in the cell, and
\begin{equation}
\mathbf{ u}_{np}(\mathbf{q},j)=\Big(\frac{\hbar}{2\rho V\omega_{\mathbf{q},j}}\Big)^{1/2}\mathbf{d}_{\mathbf{q},j}
e^{i\mathbf{q}\cdot \mathbf{x}(np)},
\label{disp}
\end{equation} 
is the amplitude of the phonon in a mode $\mathbf{q},j$ at a site (np), 
 V is the crystal volume, $\mathbf{d}_{\mathbf{q},j}$ is the phonon polarization vector, and
$\mathbf{x}(np)$ is the lattice vector. 

In calculating decoherence we assume the presence of uniform magnetic field $\mathbf{H}|| z$ with Larmor frequency $\omega_z$.  
The projection of the spin operator $\mathbf{\sigma}_z$ is conserved. Hence, 
only pure decoherence 
can affect the qubit, once it is in a superposition of up and down state, with the pure decoherence constant 
$\Gamma_{\perp}$ defining the motion of the transverse average spin:
\begin{equation}
\dot S_x = (\omega_z \times {\mathbf S})_x -\Gamma_{\perp} S_x.
\end{equation}
For the spin-phonon interaction (\ref{spinphonon}), $\Gamma_{\perp}=\Gamma_{xx}=\Gamma_{yy}$. 

Our task is to find a microscopic mechanism(s), which result in non-zero $\Gamma_{\perp}$, and to 
calculate this constant. It is important to realize that spin-phonon interaction (\ref{spinphonon}) will not lead 
to pure decoherence on its own.
The effective magnetic field that acts on transverse spin due to this interaction is time and 
coordinate-dependent. If phonons are coherent and do not decay, they would result 
in coherent modulation of the ground and excited QD Zeeman energy levels, with time-dependent phase. 
Indeed, at a given point in space, spin interaction with phonons is periodic in time.
One frequent approach to decoherence is to introduce a phenomenological 
decay/correlation time. As we shall see, the overwhelming majority of phonon decay processes do not give $\Gamma_{\perp}$. In particular, although it was suggested \cite{Golovach} 
that spin-phonon interactions give decoherence because bulk phonons enter and leave
the QD, such process cannot contribute on its own. The fact that spin-phonon interaction occurs only in a QD is accounted by 
the qubit wavefunctions defining $V_{\mathbf{q},j}$, and additional factors of phonon decay need to be considered.

To find out limitations on processes leading to pure decoherence, we calculate $\Gamma_{\perp}$,
 which is defined (see, e.g., \cite{Slichter}) by the Fourier component of the correlator of fluctuating spin-orbit 
fields at zero frequency
\begin{equation}
\Gamma_{\perp}(\omega=0)=\frac{1}{2}\int_{-\infty}^{\infty}\langle\Omega_z (0)\Omega_z(\tau)\rangle d\tau,
\label{Gamma}
\end{equation}
where angular brackets denote statistical average. The correlator in Eq.(\ref{Gamma}) is defined by the Fourier 
image of the phonon correlation function 
$C(\omega)=\frac{1}{2}\sum_{\mathbf{q}} |V_{\mathbf{q}}|^2\langle c^{\dagger}_{\mathbf{q}} (\tau) 
c_{-\mathbf{q}}(0)\rangle$+h.c., given by\cite{agd,zub}
\begin{equation}
C(\omega)=\frac{1}{\pi}\sum_{\mathbf{q}}|V_{\mathbf{q}}|^2\frac{(2N(\omega)+1) 
\Pi(\omega)}{(\omega^2-\omega_{\mathbf{q}})^2/\omega_{\mathbf{q}}^2+\Pi^2(\omega)}.
\label{corr}
\end{equation}
Here $N(\omega)$ is the number of phonons with energy $\hbar\omega$, and $\Pi(\omega)$ is the 
weighted phonon spectral function (in terms of Keldysh formalism, the imaginary 
part of retarded phonon self-energy). Real part of phonon self-energy, renormalizing phonon energy, 
does not result in decohehence and is neglected. We aim at finding 
$\Pi(\omega)$ on rigorous microscopic grounds, without introducing any phenomenological parameters of phonon decay.

The physics of dependence of $\Gamma_{\perp}$ on $\Pi(\omega=0)$ is that no energy exchange between Zeemann levels and phonons occur.
 SO hole-phonon interactions are only shaking up Zeeman levels. Physics is 
reminiscent of that of the zero-phonon line width in exciton or molecular absorption/emission 
\cite{Krivoglaz,Skinner,Gupalov}. However, spin dephasing and exciton linewidth problems differ. In the 
latter, carrier-phonon interactions give linear and quadratic  
in the phonon coordinates terms in two-level systems, and quadratic terms have been considered 
the principal cause of zero-phonon linewidth. In the former, the SO carrier-phonon term 
linear in phonon coordinates depends on $\sigma_z$, its square is spin-independent, and only the linear 
term can contribute to $\Gamma_{\perp}(\omega=0)$. Some works \cite{Skinner,Gupalov} conclude that 
interactions linear in $V_{\mathbf{q}}$ do not contribute to linewidth, or contribute only in the 
case of model localized phonons, 
rather than for realistic physical phonon/vibration modes. We now show two cases 
with real phonons contributing to relevant $\Gamma_{\perp}(\omega=0)$. 

In {\rm case I},
$\Pi^{{\rm I}} (\omega)=\omega/(2\omega_{\mathbf{q}}\tau^{{\rm I}}_{\omega_\mathbf{q}})$.
Here $1/\tau^{{\rm I}}_{\omega_\mathbf{q}}$ is the phonon decay rate at $\omega=\omega_{\mathbf{q}}$ for mechanisms giving 
$1/\tau^{{\rm I}}_{\omega}\propto \omega$. 
This is the case of Ohmic phonon spectral function, discussed in the theory of 
dissipative quantum tuneling by Leggett et all \cite{Leggett}. Here we calculate  $\Gamma_{\perp}(\omega=0)$ due to the process, 
in which phonons that interact with QD spin decay as a result of the phonon 
scattering off delocalized holes neccesarily present in the system with electrostatically confined QDs.
The {\rm case II}, in which $C(\omega=0)$ contributes to $\Gamma_{\perp}$, was never treated, and occurs when 
$
\Pi^{{\rm II}} (\omega)=\omega_{\mathbf{q}}/
(2\omega\tau^{{\rm II}}_{\mathbf{q}})
$
Here $1/\tau^{{\rm II}}_{\omega_\mathbf{q}}$ is phonon decay rate for mechanisms giving 
$1/\tau^{{\rm I}}_{\omega}\propto 1/\omega$. We find the mechanism due to phonons interacting with qubit coupled to 1D linear chain, in which phonons decay due to 
scattering of the free ends, giving the case II. 

By definition, $\Pi (\omega)$ is an odd function of $\omega$. It is easy to see that 
cases I and II give the only possible dependencies of $\Pi (\omega)$ that contribute to non-zero  
meaningful $C(\omega=0)=lim_{\omega\rightarrow 0}C(\omega)$. 
The resulting $\Gamma_{\perp}(\omega=0)$ is
\begin{eqnarray}
&\Gamma_{\perp}^{\rm I}=\sum_{\mathbf{q}}\Omega_{\mathbf{q}}^2\frac{k_BT}{\hbar\omega^3_{\mathbf{q}}\tau_{\omega_\mathbf{q}}^I}& \hspace{5mm} {\rm case}\hspace{1mm} {\rm I},
\label{Ohmic}\\
&\Gamma_{\perp}^{\rm II}=\sum_{\mathbf{q}}\Omega_{\mathbf{q}}^2\frac{k_BT}{\hbar\omega_{\mathbf{q}}}\tau_{\omega_\mathbf{q}}^{II}& \hspace{5mm} {\rm case}\hspace{1mm} {\rm II},
\label{main}
\end{eqnarray}
$T$ is the temperature, $k_B$ is the Boltzmann constant. Remarkably, $\Gamma_{\perp}\propto T$. This contrasts with
known mechanisms of spin relaxation and decoherence in QDs, but is not surprising for 
pure dephasing, which, at $\omega=0$, is similar to classical mean field fluctuations \cite{Landau}. 
We note in passing that spin-independent QD electron-phonon interaction with consequent phonon decay due to electrons in contacts 
gives {\rm case I} pure decoherence that can explain  $T^{-1}$ dependence of dephasing time in experiments \cite{marcus}. 

We now turn to microscopic calculation of $\Gamma_{\perp}$. \\
{\it Case I}: In GaAs systems, 
the piezoelectric interaction of holes with acoustic phonons that defines the last term in Eq.(\ref{holegen}) and spin-phonon interaction Eq. (\ref{spinphonon}) is given by
\begin{equation}
\hat {V}(\mathbf{r},t)= \sum_{\mathbf{q},j}\beta_{ikj}e_k e_i\mathbf{d}_{\mathbf{q},j}
e^{i\mathbf{q}\cdot \mathbf{r}-i\omega_{\mathbf{q},j}t}c^{\dagger}_{\mathbf{q},j}\sqrt{\frac{\hbar}{2\rho V\omega_{\mathbf{q},j}}}+h.c.,
\label{hph}
\end{equation}
where $\beta_{ikj}=\beta|\epsilon_{ikj}|$ is a piezoelectric constant, $\epsilon_{ikj}$ is an antisymmetric tensor, $\beta_{GaAs}=1.2\times 10^{7} $eV/cm, $e_i=q_i/q$. 
Denoting $\lambda_i=\sqrt{\hbar/m^*\omega_i} $ the sizes of the wavefunctions $\phi$ in 1D parabolic potentials, we find
 the QD orbital ground state wavefunction $\Psi\simeq\Psi_{00}+b\Psi_{11}$, where $\Psi_{nm}=\phi_n(x)\phi_m(y)$,  and 
$a= i\omega_c (\lambda_1/\lambda_2-\lambda_2/\lambda_1)/4(\omega_1+\omega_2)$. In weak magnetic field, the lowest QD states are due to Zeeman doublet of the ground orbital state. 

The hole-phonon interaction $\tilde{V}_{2}$  in Eq.(\ref{holegen}) defined by (\ref{hph}) gives $\Omega_z\propto \langle\Psi^*|e^{i\mathbf{q}\cdot \mathbf{r}}(q_xk^y-k_xq_y)|\Psi\rangle=a(q_x^2\lambda_x/\lambda_y-q_y^2\lambda_y/\lambda_x)$. It also leads to factor $A(\mathbf{q})=\sum_j|\mathbf{d}_{\mathbf{q},j}|^2$  entering $|V_{\mathbf{q}}|^2$ in Eq. (\ref{corr}).
In a simple model with isotropic speed of sound $s$, $A(\mathbf{q})=(q_x^2q_y^2+q_y^2q_z^2+q_z^2q_x^2)/q^4
$. 
 
Calculating $\Pi (\omega)$ due to interactions of phonons with holes in contacts,  we neglect renormalization of the hole-phonon vertex, like in consideration \cite{agd} of electron-phonon interactions in 3D metals. In the 2D case
\begin{equation}
\Pi ^{\rm I}_{\mathbf{q}} (\omega)= \frac{\hbar \beta^2 A(\mathbf{q}) }{2\rho \omega_q d} \int G(\mathbf{p},\epsilon)G(\mathbf{p}- \mathbf{q},\epsilon-\omega)d^2p d\epsilon
\end{equation} 
where $G(\mathbf{p},\epsilon)$, $\epsilon_F$ and $p_F$ are the Green function, Fermi energy and momentum for holes in contacts.
At small $\omega$
\begin{equation}
 \Pi^{\rm I}_{\mathbf{q}}  (\omega)= \frac{\beta^2 m_* A(\mathbf{q}) }{4\pi \hbar  d \rho\omega_q}\frac{p_F}{q}\frac{\hbar\omega}{\epsilon_F}\Theta(2p_F-q),
\label{Pi}
\end{equation}
where $\Theta(x)=1$ at $x>0$ and $\Theta(x)=0$ at $x<0$. 

Calculating the hole ground state diagonal matrix element of (\ref{hph}), evaluating Eq.(\ref{Pi}), and taking $\omega\rightarrow 0$ in Eq.(\ref{corr}), for $m_*=0.25 m_0$,  hole density in contacts $p=6\times 10^{11}$ cm$^{-2}$, $s=4\times 10^5$ cm/s, $d=100\AA$, $\lambda_x= 400\AA$, 
$\lambda_y=250\AA$, $H=0.5T$ and $T=100mK$, from Eq.(\ref{Ohmic}) we get $\Gamma_{\perp}(\omega=0)=10^4$ Hz. Thus, pure decoherence is significant in GaAs QD.  This rate exceeds 10 times the hole decoherence rate experimentally observed in InAs QD \cite{Gerardot}.  To fight the pure decoherence, one makes QD shape close to circular. We note that in SiGe/Ge/SiGe QD structure, our estimates show $\Gamma_{\perp}(\omega=0)=10^3$ Hz for similar QD sizes. With decoherence due to nuclei in such structures weaker than in GaAs, pure dephasing proposed here becomes important. We address hole qubits in Si/Ge structures elsewhere\cite{YLGHSC}. 

Phonon decay due to anharmonic processes, scattering by impurities, surfaces, or QD or QW boundaries gives $\Pi(\omega)\propto \omega^s$ with $s\ge 3$ and does not lead to case I pure dephasing. Therefore observation of pure dephasing will make it possible to separate phonon decay due to charge carriers in contacts from all other decay mechanisms.  

{\it Case II}: This setting can be realized in QD coupled to 1D linear atomic chain, where phonon decay occurs due to scattering off its free ends. 
Spin-phonon term (\ref{spinphonon}) can arise in both electron and hole QDs due to g-factor fluctuations \cite{Kim}. 
For case II in electron QDs, we consider g-factor modulation in a QD of wavefunction extent $\lambda$ by phonons of a chain of N atoms of mass M with lattice constant $a$ linked by nearest neighbor springs of spring constant $\eta$. Following \cite{Roth}, we get interaction (\ref{spinphonon}) 
$V_{q}=i(\hbar/2MN \omega_q)^{1/2}Aqe^{-q^2d^2/4}\hbar\omega_Z $, where $ \omega_{q}^2=(4\eta/M)sin^2(qa/2)$, and $A$ is a constant, which can be obtained from spin resonance experiments \cite{ESR}.
We consider ends as defects in crystal with cyclic boundary conditions. In this model,
$\Pi_q^{II}(\omega)=\frac{2\pi}{\hbar}\sum_{\mathbf{q}^{\prime}}|\langle \mathbf{u}_{\mathbf{q}^{\prime}}|T(\omega+i0)|\mathbf{u}_{\mathbf{q}}\rangle|^2\delta(\hbar\omega-\hbar\omega_{\mathbf{q}^{\prime}}).$
Here $\mathbf{u}_{\mathbf{q}}$ is given by Eq.6 in 1D case, scattering matrix $T(\omega)$ with dimensionality of a spring constant is defined by Dyson equation $T=\delta L+ \delta L G T$, $G$ is the ideal chain phonon Green function, $\delta L $ is perturbation in spring constants due to chain ends at atoms $n=1$ and  $n=0$, 
with zero $\gamma$ of spring joining them. The nonzero 
$\delta L_{nn^{\prime}}$ are $\delta L_{00}=\delta L_{11}=-\delta L_{10}=-\delta L_{01}=\eta$. Expanding $T(\omega)$ in $\mathbf{u}_{\mathbf{q}}$ as
$T_{nn^{\prime}}(\omega)= 2M^2/\hbar\sum_{q_1,q_2}u_{n}(q_1)t(q_1,q_2,\omega)u_{n^{\prime}}(q_2)$, from the solution of Dyson equation\cite{Maradudin} we find 
\begin{equation}
t(q_1,q_2,\omega^2+i0)=\frac{4\eta e^{\frac{-ia(q_1+q_2)}{2}}\sin{\frac{q_1a}{2}}\sin{\frac{q_2a}{2}}}{M\omega^2\sum_q [\omega^2-\omega_q^2+i0]^{-1}}.
\label{t}
\end{equation}
 Eq.(\ref{t}) gives $t\propto 1/\omega$ that is key to singular weighted phonon spectral function and $\Gamma_{\perp}^{\rm II}$: 
\begin{eqnarray}
&\Pi_q^{II}(\omega)=2\eta|\sin{qa/2}|/M\omega\\
&\Gamma_{\perp}^{\rm II}=4(\omega_Z/\omega_L)^2\cdot (kT/\hbar\omega_L) \cdot(A^2\hbar/M\lambda a),
\end{eqnarray}
where $\omega_L^2=4\eta/M$. Physics of $t\propto 1/\omega$ resulting in such $\Pi_q^{II}$ is clear from the electrical circuit equivalent to an atomic chain with missing spring, Fig 2c. At capacitance $C=\infty$ (due to $\gamma=0$ between ends 0 and 1), the circuit is shorted, and $\omega\rightarrow 0$ responce to finite current $J$ is zero voltage $V$
(the responce to finite $V$ is $J\propto 1/\omega$). In a chain with free ends, finite force gives displacement $\propto 1/\omega$; when phonons scatter,  $t\propto 1/\omega$.  Taking $\lambda=50A$ for a carbon QD coupled to a chain of carbon atoms with $M=2\times 10^{-23}$g, $a=2.5\AA$, 
$\omega_L=1.2\times 10^{11}$ Hz, $A=0.1$, $H=0.5$T, $T=100$ mK, we get $\Gamma_{\perp}^{\rm II}=10^7$ Hz. Such decoherence 
must be avoided. This shows, e.g., the limits on reducing bath dimensions in engineering qubits. 
Once phonons are 3D,  case II is not realized. We note here that decay rates $\Gamma_{\perp}^{\rm I}$ and $\Gamma_{\perp}^{\rm II}$ are not additive: once case II decoherence is present, $\Gamma_{\perp}^{\rm II}$ solely defines $\Gamma_{\perp}$.

In conclusion, we demonstrated that pure dephasing can considerably affect qubits in important settings. We presented two physical mechanisms 
based entirely on realistic microscopic quantum processes, without introduction of any kind of phenomenological correlation time. We also demonstrated 
that important spin-orbit interactions distinct in origin and symmetry from known Rashba and Dresselhaus terms describe physics of low-dimensional holes.  
 Support by NSF grant ECCS-0901754 is gratefully acknowledged.


\begin{thebibliography}{999}
\bibitem{Slichter} See, e.g., C.P. Slichter, Principles of magnetic resonance, Springer (1996)
\bibitem{Krivoglaz} M.A. Krivoglaz, Soviet Physics-Solid State, {\bf 6} 1340-1346 (1964)
\bibitem{Skinner} J.L. Skinner and D. Hsu, Journal of Physical Chemistry, {\bf 90} 4931 (1986)
\bibitem{Gupalov} S.V. Goupalov, R.A. Suris, P. Lavallard and D.S. Citrin, Nanotechnology, {\bf 12}, 518 (2001).
\bibitem{Arovas} D.P. Arovas and Y.B. Lyanda-Geller, Phys. Rev. B., {\bf 57} 12302 (1998)
\bibitem{Winkler} R. Winkler, Spin-Orbit Coupling Effects in Two-Dimensional Electron and Hole Systems, Springer-Verlag Berlin Heidelberg (2003)
\bibitem{finestructure} Y. Lyanda-Geller, T.L. Reinecke and M. Bayer, Phys. Rev B., {\bf 69}  161308 (2004).
 \bibitem{Lossrelax} D.V. Bulaev DV and D. Loss,
Phys. Rev. Lett.,  {\bf 95}  076805  (2005) 
\bibitem{bernevig} B. A. Bernevig and S-C Zhang, Phys. Rev. Lett. {\bf 95}, 016801 (2005).
\bibitem{Schlieman} J Schliemann and D Loss, Phys. Rev. B {\bf 71}, 085308 (2005)
 \bibitem{Ulloa} Zarea, M and Ulloa, S. E.
Phys. Rev. B {\bf 73} 165306 (2006)
\bibitem{Lin}C-X Liu, B. Zhou, S-Q Shen, and B. Zhu, Phys. Rev. B {\bf 77}, 125345 (2008) 
\bibitem{Losssr} D.V. Bulaev and D. Loss, Phys.Rev. Lett., {\bf 98}  097202  (2007)
\bibitem{Zuelicke} T. Kernreiter, M. Governale and U. Zulicke, New Journal of Physics,  {\bf 12}, 093002 (2010)
\bibitem{chesi} S. Chesi, L.P. Rokhinson, L.N. Pfeiffer, and K.W.  West, Phys Rev. Lett, {\bf 106} 236601 (2011).
\bibitem{kimSiGe} B.A. Glavin and K.W. Kim, Phys. Rev. B 71, 035321 (2005) 
\bibitem{linear} In consideration of 2D and QD ground state holes spin-orbit interactions in iii-v systems during in \cite{Arovas,Winkler,finestructure,Losssr,Ulloa,Lin,Schlieman,bernevig,Lossrelax,Zuelicke,chesi}, a term of bulk Luttinger Hamiltonian 
$\delta {\cal H}=\delta_\alpha\sum_i ( -J_i^3 \kappa_i + V_i k_i(k_i^2-k^2/3))$, where $V_z={J_z(J^2_{x}-J^2_{y})}$, $\kappa_z=k_z(k^2_{x}-k^2_{y})$, and 
other components of $V_i$ and $\kappa_i$ are obtained by cyclic permutation of indices, was not considered. In 2D and QD, this term results in linear in $\mathbf{k}$ spin-flipping term and is relevant. In 2D and QD Si-Ge systems such effect is absent, and only spin-flipping terms arising from the effects proposed in 
this letter are important \cite{YLGHSC}.
\bibitem{YLGHSC} The results for spin-flip terms using projection method 
will be given elsewhere, Y. Lyanda-Geller (unpublished). 
\bibitem{Dresselhaus} G. Dresselhaus. Phys. Rev B, {\bf 100} 580 (1955)
\bibitem{rashba1} E.I. Rashba, Soviet Physics-Solid State, {\bf 2} 1109-1122 (1960)
 \bibitem{Luttinger} J.M. Luttinger, Phys. Rev., 102 1030-1041 (1956)
\bibitem{nedoresov} S.S. Nedorezov, Sov. Phys. Solid State, 12 1814 (1971) [Fizika Tverdogo Tela, 12, 2269 (1970)] 
\bibitem{portnoi} I.A. Merkulov, V.I. Perel' and M.E.Portnoi, Zh. Eksper. Teot. Fiz 99 1202 (1991)
[Sov Phys. JETP, {\bf 72} 660 (1991)]
\bibitem{khaetsky} M.I. Dyakonov and A.V. Khaetskii, Zh. Eksp. Teor Fiz. {\bf 82} 1584 (1982)[Sov. Phys. JETP, 55 917 (1982)]
\bibitem{note} The procedure used here gives the accuracy for the 2D hole parameters, 
such as mass and SO constants, which cannot be achieved if only few excited levels  \cite{Winkler,Lossrelax,Lin} are taken into account 
by perturbation. In perturbation scheme, account of infinite series of parametrically equivalent terms 
due to coupling of the ground state with all quantized states is necessary to obtain the same results.  
\bibitem{longrange} Here we for brevity present equations for only a scalar long-range interaction of holes with bulk phonons; matrix interaction 
of holes with phonons due to deformation potential \cite{pikus} will be discussed in detail elsewhere \cite{YLGHSC}.
\bibitem{pikus}G.L. Bir and G.E. Pikus, Sov Phys. Solid State, {\bf 3} 2221 (1961)
\bibitem{LGM} Y.B. Lyanda-Geller and A.D. Mirlin, Phys. Rev. Lett., {\bf 72} 1894 (1994).
\bibitem{Sherman} E.I. Rashba and E.Y. Sherman, Phys. Lett. A {\bf 129}, 175-179  (1988).
\bibitem{Golovach} V. Golovach, A.V. Khaetskii and D.Loss, Phys. Rev. Lett. (2004)
\bibitem{agd} A.A. Abrikosov, L.P. Gorkov and I.E. Dzyaloshinski, Methods of Quantum Field Theory in Statistical Physics, Dover (1975). 
\bibitem{zub} D.N. Zubarev, Usp. Fiz Nauk, {71} 71 (1960) [Sov. Phys. Uspekhi, {\bf 3} 320 (1960).
\bibitem{Leggett} A.J. Leggett, S. Chakravarti, A.T. Dorsey,  M.P. A. Fischer, A. Garg,  and W. Zwerger, Reviews of Modern Physics, {\bf 59} 1 (1987).
\bibitem{Landau} L.D. Landau and E.M. Lifshitz, Statistical Physics. Part 1. Elsevier (1980).
\bibitem{marcus} A.G. Huibers, M. Switkes, C.M. Marcus, K. Campman and A.C. Gossard, Phys. Rev. Lett., {\bf 81} 200 (1998).
\bibitem{Gerardot} B.D. Gerardot, D. Brunner, P.A. Delgarno, P. Ohberg, S. Seidl, M. Kroner, K. Karrai, N. G. Stoltz, P.M. Petroff and R.J. Warburton,
Nature (London), {\bf 451}, 441 (2008).   
\bibitem{Kim} Y. G. Semenov and K.W. Kim, Phys. Rev {\bf 70} 085305 (2004).
\bibitem{Roth} L.M. Roth, Phys. Rev B, {\bf 118} 1534 (1960).
\bibitem{ESR} D.K. Wilson and G. Feher, Phys. Rev. B {\bf 124} 1068 (1961).
\bibitem{Maradudin} A. A. Maradudin, E. W. Montroll, G. H. Weiss, and I. P. Ipatova, Theory of Lattice Dynamics in the Harmonic Approximation (Academic, New York, 1971). 
\end{thebibliography}
\end{document}